\documentclass{optica-article}

\journal{opticajournal} 

\articletype{Research Article}

\usepackage{lineno}

\begin{document}

\title{Dissipative erasure of pulse-to-pulse dispersion imbalance in ultrafast burst-mode lasers}

\author{Mesut Laçin,\authormark{1,5} Paul Repgen,\authormark{1} Umut Aydemir,\authormark{4,5} and Fatih Ömer Ilday\authormark{1,2,3,*}}

\address{
\authormark{1}Faculty of Electrical Engineering and Information Technology, Ruhr-Universität Bochum, 44801, Germany\\
\authormark{2}Faculty of Physics and Astronomy, Ruhr-Universität Bochum, 44801, Germany\\
\authormark{3}Center for Complex Interactions, Ruhr-Universität Bochum, 44801, Germany\\
\authormark{4}Department of Electrical-Electronics Engineering, Bursa Uludağ University, Bursa, Turkey\\
\authormark{5}Optics and Photonics Engineering, Bursa Uludağ University, Bursa, Turkey
}

\email{\authormark{*}oemer.ilday@ruhr-uni-bochum.de} 


\begin{abstract*} 

Ultrafast GHz bursts enable laser-matter interactions where successive excitations collectively build sustained nonequilibrium states and efficient material interactions. However, directly generating pulse trains at GHz rates remains challenging. Alternatively, repetition-rate multipliers (RRMs) divide each oscillator pulse into equidistant replicas, but either suffer from dispersion imbalances that prevent simultaneous compression or are limited to few pulses per burst. Here, we introduce dissipative erasure, which first suppresses the differences by spectral filtering and then regenerates the optical bandwidth via nonlinear amplification without restoring the erased differences. We derive two scaling results: as the repetition rate increases geometrically with RRM stages, maximum path differences approach a finite limit; meanwhile, dispersive differences weaken quadratically with filtering. We demonstrate dissipative erasure experimentally, generating 100-$\mu$J bursts, comprising thousands of 100-fs pulses at 51.3 GHz, from a Yb-fiber amplifier. Dissipative erasure has diverse potential applications for  suppressing deterministic or stochastic differences before recovering desired optical properties.

\end{abstract*}


\section{Introduction}\label{intro}

Ultrafast laser sources delivering bursts of pulses at GHz to THz intra-burst repetition rates enable diverse laser-matter interaction regimes that are inaccessible with conventional single-pulse excitation. Their common physics is that each pulse triggers a nonlinear excitation that relaxes with a certain timescale. If subsequent pulses are incident before the excitation decays, their individual effects can build on one another nonlinearly, producing sustained nonequilibrium states or material modifications, such as ablation, that can be fundamentally different from those caused by isolated pulses. As such, bursts have found diverse applications, including precision material processing, including surface ablation~\cite{Kerse2016,Park2023} and in-volume glass modification~\cite{Lafargue2025GlassBurst}, nonlinear microscopy~\cite{burst-microscopy}, accelerators, spectroscopy, and THz generation~\cite{Buchmann2020,Khalili2024,THz-bursts}. Many of these benefits could in principle also be obtained from continuous operation at the same repetition rate, but only at prohibitively high average powers; burst operation avoids this by grouping the pulses into short bursts separated by comparatively long intervals~\cite{kalaycioglufirstburst,kalayciouglu1-mJburst,Ilday-burst-review}. 

However, generating energetic femtosecond pulses at repetition rates of tens of gigahertz or even higher remains extremely challenging. Fundamentally modelocked sources commonly have MHz repetition rates. Harmonic modelocking provides an elegant alternative, but maintaining stable, short pulses becomes increasingly difficult as the harmonic order increases~\cite{Brownian,multisoliton-burst}. Alternatives that do not rely on modelocking, such as those based on electro-optic modulation or microresonator frequency combs, have made much progress, yet they introduce other practical challenges associated with seed power, pulse energy, amplification, or system complexity. 

A particularly practical alternative to direct generation of GHz pulses is to use a modelocked source operating at a conventional repetition rate of, say, 100 MHz, and reach higher rates by dividing the pulses into equidistantly positioned replicas, using repetition-rate multipliers (RRMs). The first burst-mode laser to demonstrate this approach, to our knowledge, used a cascade of fiber-based Mach--Zehnder interferometers to reach 3.5 GHz~\cite{3.5GHzburst}. Each stage consists of an unbalanced Mach--Zehnder interferometer whose unequal arm lengths delay one copy of each input pulse by half the pulse spacing, thereby doubling the repetition rate. After $m$ cascaded stages, a single input pulse is divided into $2^m$ temporally interleaved pulse replicas. Other demonstrations used birefringence-based pulse splitting, similar to divided-pulse amplification~\cite{Zhou2007}, but, despite reaching an impressive 110 GHz, each burst was limited to only eight pulses~\cite{Wang2025}. Birefringent splitting also suffers a dispersion imbalance similar to that addressed here, although the limited scalability to longer delays keeps this effect small. Free-space delay lines have also been used~\cite{Klenke2013DividedPulse},  but scaling is perhaps even more difficult. Even THz repetition rates have been attained~\cite{THz-bursts,Stummer2025}, using the Vernier effect, but the demonstration was also limited to a small number of pulses per burst. Other multiplication approaches include frequency-comb filtering using external cavities~\cite{Kirchner2009} that require interferometric stabilization and incur substantial optical power losses, and the temporal Talbot effect using dispersive elements~\cite{Longhi2000}, which demands precisely controlled dispersion and sufficient coherence between pulses. Active fiber loops offer another approach, using AOM-controlled pulse circulation with dispersion and loss compensation~\cite{Bartulevicius2020,Bartulevicius2022,ActiveFiberLoop2024}, but require active control and precise compensation of dispersion and loss during each round trip.

Overall, RRMs employing only passive fiber components are extremely robust, comparatively inexpensive, and are scalable to hundreds or even thousands of pulses,~\cite{LacinEurophoton2024}. However, because the temporal delays require unequal fiber lengths in the interferometer arms, each pulse replica exiting the RRM chain has accumulated a different amount of chromatic dispersion. Since any dispersion compensation would apply equally to all of them, they cannot be simultaneously dechirped to their shortest durations after burst-mode amplification. Furthermore, amplification in fibers is often strongly nonlinear, such as in the self-similar (similariton)~\cite{similariton-amp} or gain-managed amplification regimes~\cite{Sidorenko2019,Sidorenko2020}. Although these regimes show attractor-like evolution, finite amplifier lengths severely limit the extent to which differences in the seed pulses can be suppressed, so the pulses follow different nonlinear evolution trajectories, further aggravating the problem. In theory, this dispersion imbalance could be avoided by using narrowband seed pulses from the beginning. In practice, the large energy division and insertion losses of many RRMs require a certain pulse energy before the RRM chain, and it is highly desirable to start with broadband pulses, so they can be chirped to limit the nonlinear phase shift before the RRMs, which, if uncontrolled, would be even more detrimental to the burst-mode amplification.

\begin{figure}
\centering
\includegraphics[width=0.85\textwidth]{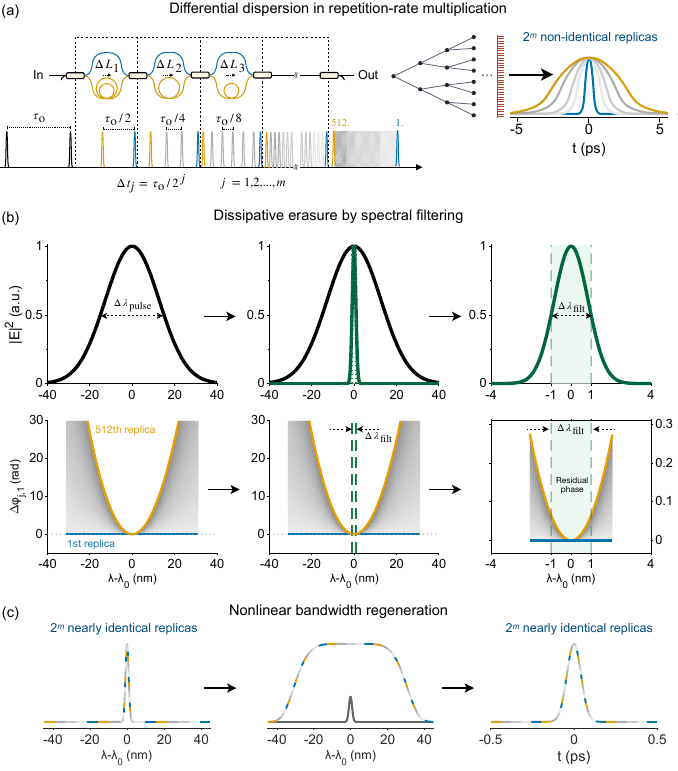}

\caption{Concept of dissipative erasure of pulse-to-pulse dispersion imbalance followed by nonlinear bandwidth regeneration. (a) $m$ cascaded repetition-rate multipliers (RRMs) increase the repetition rate by $2^m$ by splitting and delaying the input pulses. Because the resulting pulse replicas propagate through different fiber lengths, they accumulate different amounts of dispersion and emerge from the RRM chain with different chirps. Blue and orange denote the first and last replicas, respectively, while intermediate replicas are shown in gray. (b) Restricting the original spectral bandwidth, $\Delta\lambda_{\rm pulse}$, to the narrower filter bandwidth, $\Delta\lambda_{\rm filt}$, selects a region over which the differential dispersive phase is small. The upper panels show the input spectrum (black) and filtered spectrum (green). The lower panels show spectral phase differences relative to the first replica, $\Delta\phi_{j,1}=\phi_j-\phi_1$. Darker gray shades indicate larger differential accumulated dispersion relative to the first replica. The pale-green shading and dashed boundaries indicate the filter bandwidth. (c) Subsequent nonlinear amplification regenerates the bandwidth required for femtosecond pulses without regenerating the erased pulse-to-pulse differences. The gray curve in the middle panel represents the filtered seed, and the right panel illustrates the compressed output pulses. Colored segments represent the replica ensemble. }

\label{fig:concept}
\end{figure}

Here, we theoretically propose and experimentally demonstrate a dissipative-regenerative scheme that erases the dispersion imbalance between the pulse replicas, as illustrated conceptually in Fig.~\ref{fig:concept}. Rather than attempting to compensate the different propagation histories individually, strong spectral filtering dissipatively projects the replicas toward a common pulse shape by restricting their bandwidth to a region over which the differential dispersive phase becomes sufficiently small. Subsequent nonlinear amplification regenerates the broad optical bandwidth required for femtosecond pulse generation without recreating the erased pulse-to-pulse differences. Fundamentally, the erasure step brings the initially different pulses into the basin of attraction of the same nonlinear state, which nonlinear amplification alone cannot achieve within a finite amplifier length. We demonstrate the scheme in a Yb-fiber system in which nine cascaded RRMs raise a 100.3-MHz seed repetition rate to 51.3~GHz. For 100-ns, 210-ns, and 400-ns burst settings, the amplifier generates approximately 100-fs pulses at an average output power of approximately 100~W and burst energies of approximately 100~$\mu$J after the grating compressor.

\section{Operating Principle}

\subsection{Dispersion accumulation in repetition-rate multipliers}

As illustrated in Fig.~\ref{fig:concept}(a), the delay introduced by each RRM stage equals half the pulse spacing at its input. Denoting the oscillator period by $\tau_{\rm o}$, the fiber length difference introduced by the first stage is
\begin{equation}
\Delta L_1=\frac{v_{\rm g}\tau_{\rm o}}{2},
\end{equation}
where $v_{\rm g}$ is the group velocity in the fiber.

Each successive stage requires half the differential delay of the preceding stage. Since each pulse replica follows one of the two arms at each RRM stage, the additional fiber lengths are given simply by binary combinations of these delays. After $m$ stages, the $N=2^m$ replicas, ordered by their delays, have path length differences
\begin{equation}
\Delta L_j
=v_{\rm g}\tau_{\rm o}\sum_{k=1}^{m}\frac{b_k}{2^k}
=\frac{j-1}{N}v_{\rm g}\tau_{\rm o},
\qquad j=1,\ldots,N,
\label{distribution}
\end{equation}
where $b_k$ is either 0, or 1, depending on whether the short path or the long path is taken at the $k^{\rm th}$ RRM, respectively. Thus, these differences form a uniform discrete distribution, with spacing $v_{\rm g}\tau_{\rm o}/N$, with the maximum,
\begin{equation}
\Delta L_{\rm max}
=\sum_{k=1}^{m}\frac{v_{\rm g}\tau_{\rm o}}{2^k}
=v_{\rm g}\tau_{\rm o}(1-2^{-m}).
\end{equation}

This yields the first advantageous scaling, namely that the number of replicas, thus the final repetition rate, increases geometrically with $m$, while the maximum path difference quickly converges to the constant value, $\Delta L_{\rm max}\simeq v_{\rm g}\tau_{\rm o}$, set by the seed laser's repetition rate.  

For a seed repetition rate of 100~MHz, the maximum path length difference is roughly 2~m. The resulting dispersion difference causes a group-delay exceeding a picosecond across a bandwidth broad enough to support sub-100-fs pulses, revealing the severity of the problem, despite the fortunate scaling. 

\subsection{Dissipative erasure of dispersion imbalance}

Our approach, depicted in Fig.~\ref{fig:concept}(b,c), does not attempt to compensate the accumulated dispersion of individual pulse replicas. Instead, the pulse train is first passed through a narrowband spectral filter before the final amplification stage. The filter reduces the optical bandwidth enough to render the dispersive delay differences small compared to their transform-limited durations. For a 100-MHz initial repetition rate, about 2~nm suffices, corresponding to a transform-limited duration about 0.8~ps. Then, the differences among the pulse replicas become small relative to their durations, making all the pulses nearly identical. It is in this sense that spectral filtering acts as a dissipative eraser of the pulse-to-pulse differences.

Following spectral filtering, the homogenized pulse train is amplified in a strongly nonlinear Yb-fiber amplifier, where the optical bandwidth required for femtosecond pulse compression is generated, while the nearly identical initial conditions ensure that all pulse replicas undergo nearly identical nonlinear evolution. In fact, the amplifier does not merely preserve, but also reinforces this near-identity: in the self-similar regime, the asymptotic solution is an attractor whose duration, chirp and spectral width are set by the pulse energy and the fiber parameters rather than by the seed~\cite{similariton-amp,Finot2004}, so that replicas entering with slightly different chirps converge toward the same pulse shape. A similar convergence can occur in the gain-managed amplification regime~\cite{Sidorenko2019} and in laser cavities~\cite{Lu2024MamyshevGMN}. Mamyshev regenerators can, naturally, support attraction toward the same pulse state~\cite{MamyshevCascade2008}. Within finite amplifier lengths, however, convergence to nearly identical output states requires sufficiently similar initial pulses.

The effectiveness of dissipative erasure can be quantified through a simple estimate. For a bandwidth of $\Delta\lambda$, the maximum pulse-to-pulse difference in group delay is
\begin{equation}
\Delta\tau\simeq |D|\,v_{\rm g}\tau_{\rm o}\,\Delta\lambda.
\end{equation}
Following filtering, the transform-limited pulse duration is approximately
\begin{equation}
\tau_{\rm lim}=\frac{\sigma \lambda_0^2}{c\,\Delta\lambda_{\rm filt}},
\end{equation}
where $\sigma=\tau_{\rm lim}\Delta\nu$ is the time-bandwidth product, a dimensionless constant that depends on the pulse shape (0.44 for Gaussian). Then, we can estimate the maximum relative pulse duration after dispersion compensation as
\begin{equation}
\frac{\tau_{\rm max}^{(\rm filt)}}{\tau_{\rm lim}} \lesssim 1+\frac{\Delta\tau}{\tau_{\rm lim}}
= 1 + \frac{c^2|D|\tau_{\rm o}}{n_{\rm g}\sigma\lambda_0^2}\Delta\lambda_{\rm filt}^2,
\end{equation}
where $n_{\rm g}=c/v_{\rm g}$ is the group index. Then, it is easy to see that the dispersion imbalance becomes negligible when
\begin{equation}
\eta \equiv  \frac{c^2|D|\tau_{\rm o}}{n_{\rm g}\sigma \lambda_0^2}\Delta\lambda_{\rm filt}^2\ll1.
\label{criterion}
\end{equation}
This yields the second fortunate scaling, which is the quadratic dependence of the importance of the dispersion imbalance on the filter bandwidth. Narrowing the spectrum from a bandwidth supporting approximately 100-fs pulses to approximately 2 nm reduces the impact of the dispersion imbalance by two orders of magnitude. For the experimental setup reported below, with $\tau_{\rm o} = 10$~ns, $n_{\rm g}=1.47$, $D=39$~ps/(nm.km), and $\lambda_0=1060$~nm, we obtain $\eta = 0.19$ for a 2-nm Gaussian filter. This corresponds to a conservative estimate of the maximum relative pulse-duration increase of less than 20\%. Since the path lengths are uniformly distributed, the corresponding average estimate is less than 10\%. By contrast, retaining a 30-nm bandwidth through the RRM chain would mean $\eta \simeq 43$, completely precluding compression to nearly transform-limited femtosecond pulses for much of the burst. The filter bandwidth should be sufficiently narrow to suppress the dispersion imbalance, but no narrower than necessary, since this would increase the losses and make bandwidth regeneration more demanding.

\subsection{Numerical demonstration of dissipative erasure}

We next numerically confirm that spectral filtering suppresses the dispersion imbalance sufficiently for pulses with different propagation histories to undergo virtually identical nonlinear evolution in the subsequent amplifier. Pulse propagation was simulated using the generalized nonlinear Schr\"odinger equation solved by the split-step Fourier method~\cite{Senel2023UPP}, including dispersion, Kerr nonlinearity, and saturable gain with a Gaussian gain spectrum of 40-nm bandwidth centered at 1050~nm. We deliberately choose generic parameters and a simple single-stage amplification, rather than model the multi-stage experimental setup, to illustrate dissipative erasure. 

Each simulation starts from a transform-limited 100-fs Gaussian pulse centered at 1040~nm, chirped to 1~ps. The two replicas differ only in an additional 2~m of PM980 fiber traversed by $j=512$, corresponding to the maximum path difference in the RRM chain when starting from a repetition rate of 100 MHz. Intermediate pulses have intermediate path lengths and evolutions. The control case is without the use of spectral filtering. In those simulations, the chirped pulses with an energy of 50~pJ enter a 5-m-long, singlemode Yb-fiber amplifier, and reach approximately 1.5 nJ pulses. Each replica is propagated independently, and we do not consider the effect of potential gain-mediated coupling between pulses. Due to the differences in the nonlinear spectral broadening, and its impact on the amplification, the amplified pulse energies vary by up to 10\%. Both replicas are then compressed with an identical grating-compressor setting, optimized for $j=1$. In the case of dissipative erasure, a Gaussian-shaped spectral filter, also centered at 1040 nm, with a full width at half maximum (FWHM) of 2 nm, is inserted before the amplifier. The amplifier powers are adjusted to yield the same pulse energies for the first pulses. The inclusion or omission of the filter and an extra length of 2 m of fiber are the only other differences between the simulated results. 

\begin{figure}[htbp]
\centering
\includegraphics[width=0.8\textwidth]{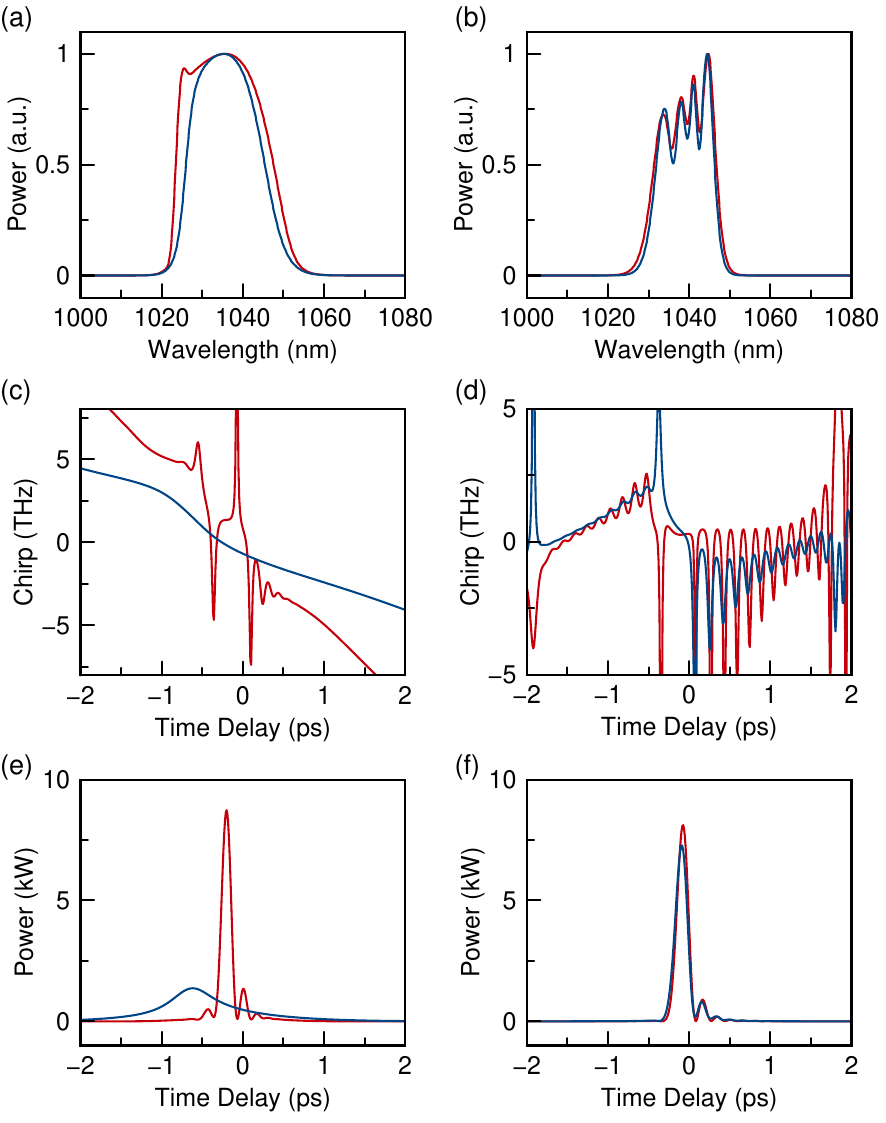}
\caption{Numerical demonstration of dissipative erasure. The first ($j = 1$, red) and last ($j = 512$, blue) replicas, which traverse the shortest and longest paths through the nine-stage RRM chain, are compared without (left column) and with (right column) dissipative erasure through spectral filtering. (a), (b) Output spectra, (c), (d) temporal chirp, and (e), (f) temporal power profiles after dechirping with the compressor optimized for first pulse. Without dissipative erasure, the first and last pulse widths differ by almost an order of magnitude, whereas the pulses are nearly identical with dissipative erasure.}
\label{fig:Erasure_numerical}
\end{figure}

Without spectral filtering [Fig.~\ref{fig:Erasure_numerical}, left column], the system operates as a chirped-pulse amplifier (CPA)~\cite{Strickland1985} with substantial nonlinear evolution during amplification~\cite{Zhou2005,Chong2007}. Because of their different initial chirps, the two replicas follow different nonlinear trajectories and develop different temporal and spectral profiles. They compress to 131~fs with a bandwidth of 24 nm and 767~fs with a bandwidth of 19 nm, when the compressor is optimized for the first pulse. Even with the compressor optimized separately for each, they reach approximately 131~fs and 154~fs: the replicas differ not only by a linear chirp offset, but also in their nonlinear evolution.

With spectral filtering [Fig.~\ref{fig:Erasure_numerical}, right column], the two replicas enter the amplifier in nearly identical states and follow nearly the same nonlinear evolution toward the same attractor state with parabolic chirped temporal profiles and nearly identical 15 nm-wide spectra. They compress to 164 and 169~fs, identical within a few femtoseconds of each other. Even optimizing the compressor individually does not yield shorter pulses, demonstrating nonlinear bandwidth regeneration without reintroducing the differences arising from their propagation histories.

\section{Experimental demonstration}

\subsection{Laser architecture and repetition-rate multiplication}

\begin{figure}[htbp]
    \centering
    \includegraphics[width=\linewidth]{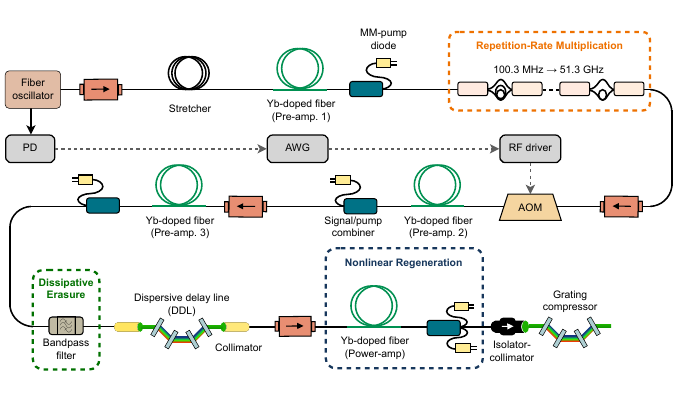}
    \caption{Schematic of the experimental burst-mode Yb-fiber laser system, highlighting repetition-rate multiplication, dissipative erasure, and nonlinear bandwidth regeneration. An acousto-optic modulator (AOM) is used to form the burst by gating the pulses and to pre-shape the burst envelope to compensate for gain depletion during amplification. After further preamplification, a narrowband PM bandpass filter suppresses the pulse-to-pulse dispersion imbalance introduced by the repetition-rate multiplier (RRM) chain through dissipative spectral filtering. A grating-based dispersive delay line (DDL) adjusts the pulse chirp before strongly nonlinear amplification in the final amplifier, where the optical bandwidth is regenerated. The pulses are then dechirped using a free-space grating compressor.}

    \label{fig:experimental_setup}
\end{figure}

Figure~\ref{fig:experimental_setup} illustrates the experimental burst-mode Yb-fiber laser system, highlighting three principal stages: repetition-rate multiplication, dissipative erasure, and nonlinear bandwidth regeneration~\cite{LacinEurophoton2024}. A home-built modelocked Yb-fiber oscillator generates pulses at a repetition rate of 100.3~MHz, centered at 1044~nm with a spectral bandwidth of 31~nm. The pulses are temporally stretched in 80~m of PM980 fiber and amplified in the first preamplifier (Pre-Amp.~1) before entering the repetition-rate multiplication stage. All amplifier stages use polarization-maintaining and singlemode fibers.

The repetition rate of the seed pulses is increased using nine cascaded fiberized RRMs, producing $N=2^9=512$ pulse replicas per oscillator pulse, reaching a repetition rate of 51.3~GHz with $\tau_{\rm r}=\tau_{\rm o}/N=19.5$~ps. Burst operation is imposed by an acousto-optic modulator (AOM) positioned after the RRMs. A photodiode (PD) detects a fraction of the oscillator output and provides an electrical synchronization signal to an arbitrary waveform generator (AWG). The AWG controls the AOM through an RF driver, enabling synchronized burst selection with adjustable duration and repetition rate, as well as programmable envelope preshaping to compensate for gain depletion during subsequent amplification. The burst repetition rate was set to 920 kHz for the present experiments.

After the AOM, the pulse train is amplified in two preamplifiers (Pre-Amp.~2 and Pre-Amp.~3), before spectral filtering. A grating-based dispersive delay line (DDL) adjusts the pulse chirp before the final power amplifier. The final amplifier comprises 3.5~m of Y1200 10/125 PM-DC gain fiber. The regenerated broadband pulses are finally compressed using a Treacy-type grating compressor.

\subsection{Experimental realization of dissipative erasure}

\begin{figure}[htbp]
    \centering
    \includegraphics[width=0.85\linewidth]{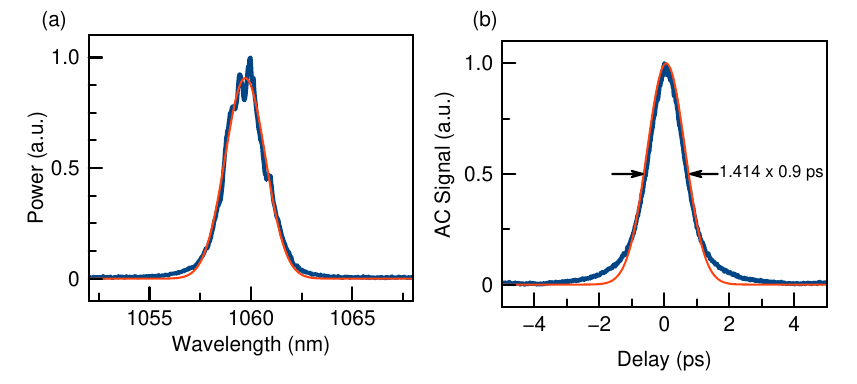}
     \caption{Experimental preparation of the homogenized pulse train before nonlinear amplification. (a) Measured seed spectrum after spectral filtering, showing an approximately Gaussian profile with a FWHM bandwidth of about 2~nm. (b) Autocorrelation trace after the dispersive delay line, with a Gaussian fit, indicating a pulse duration of approximately 0.9 ps, close to the transform-limited duration of 0.82 ps. }
    \label{fig:filtering_exp}
\end{figure}

Following the RRMs and subsequent preamplification, the pulse train is spectrally filtered using a narrowband filter with a nearly Gaussian transmission profile. The filter reduces the spectral bandwidth to approximately 2~nm. Using a grating-based DDL, we maximally compressed the filtered pulse train. Figure~\ref{fig:filtering_exp} shows the resulting spectrum and intensity autocorrelation, which necessarily average over the entire burst and cannot resolve the individual pulse replicas. The pulse width inferred from the autocorrelation measurement, assuming a Gaussian pulse shape, is 0.9~ps, which closely matches the 0.82-ps transform-limited duration calculated from the measured spectral bandwidth. Importantly, the differential dispersion among the replicas has a known distribution given by Eq.~\ref{distribution} with a strict upper bound, which precludes arbitrarily large deviations in the chirp of individual pulses. Since residual pulse-to-pulse chirp differences can only increase the duration of the affected pulses above their transform limit, this result effectively proves that such differences have become negligible compared to the pulse duration across the burst.

\subsection{Nonlinear amplification and pulse compression}

\begin{figure}[htbp]
    \centering
    \includegraphics[width=0.8\linewidth]{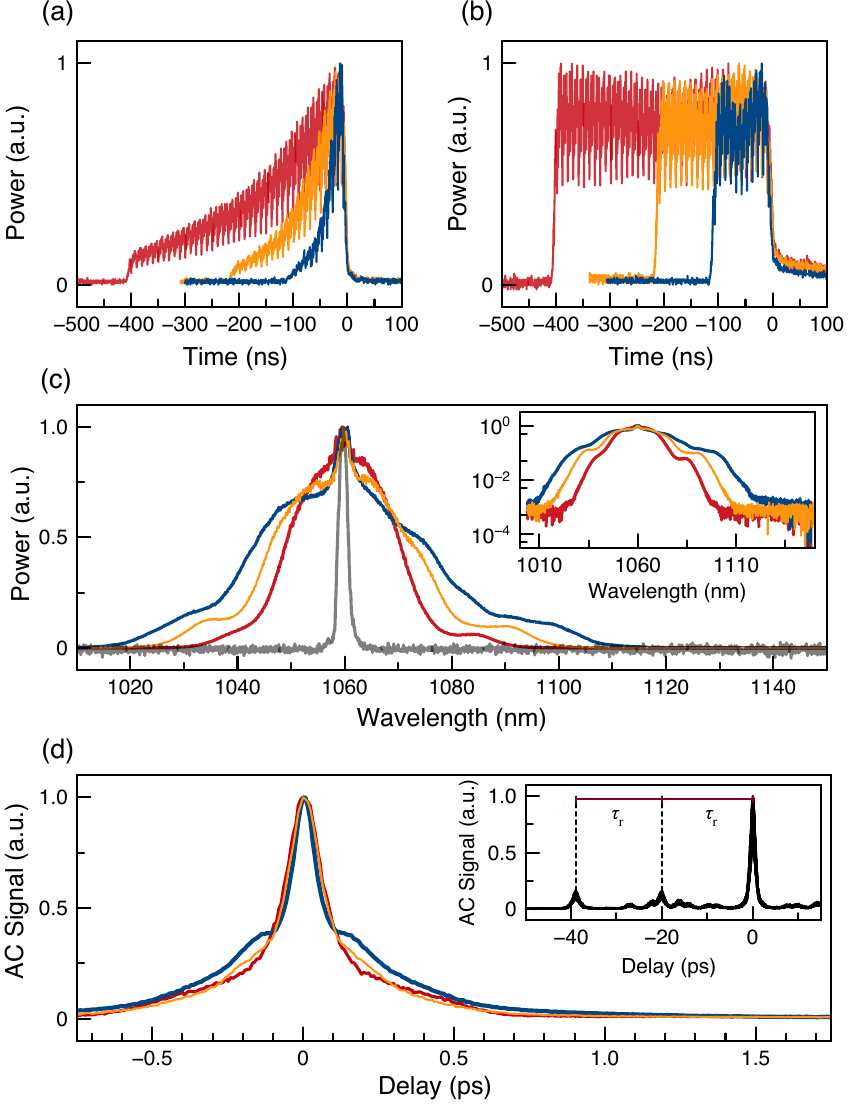}
    \caption{Final amplification and compressed-pulse characterization for 100-ns (dark blue), 210-ns (orange), and 400-ns (red) AOM-gate settings at a burst repetition rate of 920~kHz. (a) Pre-shaped burst envelopes measured at the output of the AOM following the RRM chain. (b) Burst envelopes measured after the main amplifier. (c) Main-amplifier output spectra on a linear scale, with the inset showing the spectra on a logarithmic scale. The gray curve is the filtered seed. (d) Intensity autocorrelations measured after the grating compressor. The inferred pulse durations, assuming Gaussian profiles, are 93~fs, 108~fs, and 115~fs for the 100-ns, 210-ns, and 400-ns settings, respectively. The inset shows a longer-range autocorrelation trace, demonstrating the intra-burst pulse spacing of $\tau_{\rm r}\simeq19.5$~ps, corresponding to 51.3~GHz. }
    \label{fig:final_amplification}
\end{figure}

The 0.9-ps pulses are sufficiently short compared to the intra-burst spacing of 19.5 ps, and remain temporally separated in the final Yb-fiber power amplifier, which is operated in a strongly nonlinear regime to regenerate the optical bandwidth removed during spectral filtering. Their energy at this stage is approximately 0.5-1~nJ. Figure~\ref{fig:final_amplification} presents results for three burst durations, 100~ns, 210~ns, and 400~ns, at a burst repetition rate of 920~kHz. These correspond to approximately 5,100, 10,800, and 20,500 pulses per burst, respectively. Because of the finite rise and fall times of the AOM, these burst durations and pulse counts are approximate. The AOM imposes rising burst envelopes to compensate for gain depletion [Fig.~\ref{fig:final_amplification}(a)]. This results in comparatively uniform envelopes after amplification [Fig.~\ref{fig:final_amplification}(b)], except for the modulation with period $\tau_{\rm o}\simeq10$~ns visible in both envelopes, arising from deviations of the nominally 50:50 couplers in the RRMs from equal splitting. The sensitivity of burst amplitude uniformity to coupler splitting ratios and unequal splice losses has also been reported in a cascaded fiber repetition-rate multiplier~\cite{ChenTunableBurst}.

The amplified burst energies are approximately 153, 157, and 161~$\mu$J for the three configurations, corresponding to mean pulse energies of approximately 30, 15, and 8~nJ, respectively. Figure~\ref{fig:final_amplification}(c) shows that the filtered seed is broadened to 30, 25, and 21~nm for the 100-ns, 210-ns, and 400-ns settings, respectively. After transmission through the grating compressor, which has a throughput of 71\%, the amplifier system delivers approximately 100~W of average power and burst energies of 105-111~$\mu$J. Following optimization of the grating compressor for each configuration, we infer average pulse durations of 93, 108, and 115~fs from the intensity autocorrelations, assuming a Gaussian deconvolution factor [Fig.~\ref{fig:final_amplification}(d)]. The corresponding time--bandwidth products are approximately 0.74, 0.73, and 0.63, respectively. Although above the transform limit, these values are within the expected range for strongly nonlinear fiber amplifiers. 

These results demonstrate the simultaneous generation of 5,000-20,000 pulses of $\sim 100$~fs duration at 51.3~GHz, with more than 100~$\mu$J of energy per burst after compression. This combination of high intra-burst repetition rate, large pulse count, ultrashort pulse duration, and substantial burst energy is unique to our knowledge among reported GHz-THz burst sources.

\section{Discussion and Conclusion}\label{conclusion}

We have demonstrated a dissipative-regenerative approach for overcoming the pulse-to-pulse dispersion imbalance inherent to fiberized repetition-rate multiplication. Narrowband spectral filtering suppresses the consequences of unequal propagation histories, allowing common dispersion compensation to produce a nearly transform-limited pulse train. We derived a simple metric (Eq.~\ref{criterion}) for choosing the filter bandwidth. Subsequent nonlinear amplification regenerates the bandwidth required for femtosecond pulse generation without regenerating the erased pulse-to-pulse differences. 

Experimentally, nine cascaded RRMs multiply the repetition rate of a 100-MHz oscillator by 512, reaching 51.3~GHz. Spectral filtering to approximately 2~nm followed by common dispersion compensation produces nearly transform-limited pulses with strongly suppressed pulse-to-pulse differences. Nonlinear amplification regenerates 20--30~nm of spectral bandwidth, and following dechirping, the system delivers 105--111~$\mu$J bursts containing 5,000--20,000 pulses of approximately 100~fs duration, at a burst repetition rate of 920~kHz.

Our approach is built on two favorable scaling relationships: first, even though the number of pulse replicas, and thus repetition rate, increases geometrically as $2^m$, the maximum differential propagation length, hence the significance of dispersion imbalance, converges to the finite limit $v_{\rm g}\tau_{\rm o}$, set by the seed laser's repetition rate. Second, the impact of the dispersion imbalance decreases quadratically with filter bandwidth. Consequently, increasing the repetition rate does not require narrower filtering, making our approach especially attractive for generating bursts containing large numbers of tightly spaced pulses.

There is an additional source of pulse-to-pulse nonuniformity at high burst energies due to gain depletion across the burst, which is compensated by preshaping the burst envelope before the power amplifier. At sufficiently strong gain depletion and strong nonlinearity, the pulses can no longer be regarded as independently amplified, since gain depletion intertwines their nonlinear evolution. This is already partially the case here. The influence of collective gain dynamics on pulse amplification within a burst was previously reported for gain-managed nonlinear amplification~\cite{maghsoudi50fsPulseBursts2024}. However, such collective amplification dynamics are beyond the scope of the present work and will be discussed in detail elsewhere.

More broadly, the method illustrates a useful alternative to direct compensation of deterministic or even stochastic imperfections in ultrafast optical systems. The key is to dissipatively project initially different pulses into the basin of attraction of the same nonlinear state, rather than attempting to compensate their distinct histories individually. Fundamentally, any erasure of information is necessarily dissipative~\cite{Landauer1961}: a lossless linear operation preserves their differences, and only a lossy one can make them indistinguishable. Here, this is accomplished by spectral filtering: spectral information is discarded, and with it the information corresponding to the differential propagation histories. Subsequent nonlinear amplification is itself dissipative and restores the desired optical bandwidth as the pulses evolve towards the common attractor, without reconstructing the erased information. This differs from other filtering-based discrimination schemes, such as Mamyshev regeneration~\cite{Mamyshev1998}. In our case, the aim is to homogenize all pulses by launching them toward the same attractor state, rather than selecting which ones survive. This dissipative-regenerative strategy may find applications beyond burst-mode lasers, particularly in nonlinear photonic information processing~\cite{Manuylovich2025,Calado2026}, where unwanted variations could be erased and desirable optical properties regenerated without sacrificing relevant information.

\begin{backmatter}
\bmsection{Funding}
This work was supported by the Alexander von Humboldt Foundation through an Alexander von Humboldt Professorship awarded to F. Ö. Ilday and the European Research Council (ERC) under the European Union's Horizon Europe research and innovation programme (grant agreement no. 101055055, ERC Advanced Grant UNILASE).

\bmsection{Acknowledgment}
The authors thank Aladin \c{S}ura for help with the seed oscillator, \c{C}a\u{g}r{\i} \c{S}enel for the simulation code, Amirhossein Maghsoudi for his assistance in improving the thermal contact of the gain fiber, and Robin Marjoribanks for first pointing out the dispersion-imbalance problem.

\bmsection{Disclosures}
The authors declare no conflicts of interest.

\bmsection{Data Availability Statement}
Data underlying the results presented in this paper are not publicly available at this time but may be obtained from the authors upon reasonable request.


\end{backmatter}


\bibliography{Dissipative_erasure}

\end{document}